\documentclass[twocolumn,showpacs,amsmath,amssymb,superscriptaddress]{revtex4}
\usepackage{dcolumn}
\usepackage{bm}
\usepackage{amsmath}
\usepackage{amsfonts}
\usepackage{graphics}
\usepackage{graphicx}

\begin{document}

\title{Scale invariant properties of public debt growth}

\author{Alexander M. Petersen}
\affiliation{Center for Polymer Studies and Department of Physics, Boston University, Boston, Massachusetts 02215, USA}
\author{Boris Podobnik}
\affiliation{Faculty of Civil Engineering, University of Rijeka, 51000 Rijeka, Croatia}
\affiliation{Zagreb School of Economics and Management, 10000  Zagreb, Croatia}
\affiliation{Center for Polymer Studies and Department of Physics, Boston University, Boston, Massachusetts 02215, USA}
\author{Davor Horvatic}
\affiliation{ Faculty of Science, University of Zagreb, 10000  Zagreb, Croatia }
\author{H. Eugene Stanley}
\affiliation{Center for Polymer Studies and Department of Physics, Boston University, Boston, Massachusetts 02215, USA}

\pacs{89.75.Da, 89.90.+n}


\begin{abstract}
Public debt is one of the important economic variables that quantitatively describes a nation's economy.
Because bankruptcy is a risk faced even by institutions as large as governments (e.g. Iceland),  national
debt should be strictly controlled with respect to national wealth. Also, the problem of eliminating extreme poverty in
the 
world is closely connected to the study of extremely poor debtor nations. We analyze the time evolution of national
public debt and find ``convergence":  initially less-indebted countries increase their debt more quickly than
initially more-indebted countries. We also analyze the public debt-to-GDP ratio ${\cal R}$, a proxy for default risk,
and approximate the probability density function $P({\cal R})$ with a Gamma distribution, which can be used to establish
 thresholds for sustainable debt. We also observe ``convergence" in ${\cal R}$: countries with  initially small  ${\cal R}$ increase their ${\cal R}$ more quickly than countries with initially
large  ${\cal R}$.  The scaling relationships for debt and  ${\cal R}$ have practical applications,  e.g. 
  the Maastricht Treaty requires members of the European Monetary Union to
  maintain  ${\cal R} < 0.6$.
  \end{abstract}

\maketitle
\section{Introduction}
  Just as an individual is expected to control his/her debt to asset ratio, so is a  
government expected to control its national debt as a function of  the 
 country's wealth, measured e.g. by its gross domestic product (GDP). 
In a dynamic global economy, excessive borrowing cannot persist indefinitely, as creditors are bound to call in large
loans. While a country suffers  financial problems when its GDP does not 
increase fast enough, even more serious trouble begins when its debt increases faster than its GDP.
 While national GDP has been the topic of many studies on economic growth \cite{barro91, Levine92, 
Durlauf96,barrobook}, the empirical analysis of public debt has lagged due to lack of comprehensive data.

Large sets of public debt data, dating back several decades, and ranging from poor to rich countries, 
have recently become available. 
Here we use  concepts of statistical physics to analyze public debt data for a wide cross-section of economies including underdeveloped,
developing, and 
developed countries.  
The total public debt data, along with total GDP data, are available at the {\it Inter-American Development Bank\/}
\cite{data0}, and are compiled and analyzed in Refs. \cite{JP06a,JP06b}. Population data are available 
by the {\it World Bank}, and can be reconstructed through GDP and {\it per capita\/} GDP data compiled in Ref. \cite{data1}. We deflate all
{\it USD} amounts to in units of the {\it USD} in the year 2000  \cite{data2}. In our 
analysis, we compare public debt only within the same country, in order to avoid any differences in the theoretical and
practical 
definition of debt and the reporting of debt by various countries, an issue pointed out in Refs. \cite{JP06a,JP06b,
kotlikoff}. Our results are robust with respect to mis-reporting and ambiguous definitions of 
public debt \cite{kotlikoff}. In Fig. \ref{DbyGExamples} we plot the debt-to-GDP ratio ${\cal R}$ for
many countries, grouped in panels (a-d) by common historical, geographical, and financial factors. In Fig.
\ref{DbyGExamples}(d) we plot the average debt-to-GDP ratio for three subgroups corresponding to {\it World
Bank} Income Group (IG) classifications, and observe relatively high levels of ${\cal R}$ among the poorest
countries.   

Economic growth theories predict that  GDP should ``converge" towards equality, with wealthy 
countries experiencing smaller relative growth rates than poor countries. However, the opposite has been found for
economic  wealth
data 
\cite{barromartin91, barromartin92,  martin96}. So we address the question, what are the growth dynamics for public 
debt? To answer this question, we analyze a comprehensive database of national public debt and GDP to investigate 
the dynamics of  debt growth and  growth in ${\cal R}$.

With the current global credit crunch, and several notable recent national defaults, it is important to address
sustainable public debt, defined as the amount of debt where the receiving country is capable of meeting its current and
future debt obligations
\cite{reinhart,kraay}. The total current government debt $D(t)$ increases from last year's debt $D(t-1)$ partially due
to interest payments on the debt $D(t-1)$ at interest rate $I_{D}(t)$, and partially because of the current primary
deficit, defined as the difference between spending $S(t+1)$ and taxes $T(t+1)$ \cite{blanchard85}. Thus
\begin{equation}
D(t) = [1 + I_{D}(t-1)] D(t-1) + [S(t) - T(t)] \ .
\end{equation}
We consider three possible scenarios for public debt growth dynamics:

\begin{itemize}
\item[{(i)}] Growth rates of the country debt do not depend on the initial
  debt level.

\item[{(ii)}]  A more indebted country has a {\it larger} 
debt growth rate than a less indebted country, so that relative differences
  between debt across countries increases over time ({\it divergence}).

\item[{(iii)}] A more indebted country has a {\it smaller} 
debt growth rate  than a less indebted country, so that relative differences
  between debt across countries decreases over time ({\it convergence}).
\end{itemize} 
These three scenarios  have different implications for investors, who will only accept government debt up to some
ceiling.  Hence, one would expect that more indebted countries would increase their debt more slowly than less indebted
countries.

\section{Empirical results} To ascertain which of the three debt scenarios is better supported by empirical facts, we
define for country $i$ the annualized logarithmic growth rate of {\it per capita} initial debt $d_{i}(t)$ between years $t$ and $t+\Delta t$
\begin{equation}
r_i(t,t+\Delta t)\equiv \frac{\log[d_i(t+\Delta t)/d_i(t)]}{\Delta t} \ .
\end{equation}
  We compare $r_i(t,t+\Delta t)$ 
to $d_i(t)$, assuming $r_i(t,t+\Delta t)$ depends on debt size by
\begin{equation} 
r_i(t,t+\Delta t)\cong\alpha-\beta ~\log[d_i(t)] \ .  
\label{regression}
\end{equation} 
The functional form of Eq.~(\ref{regression}) can also be expressed as
\begin{equation}
\log[ d_i(t+\Delta t)]= \alpha ~\Delta t +(1-\beta\Delta t) \log[d_i(t)]
\label{loglogregress}
\end{equation}
If $\beta > 0$, there is  convergence  in  {\it per
  capita\/} debt data across countries, since initially more indebted economies tend to increase their debt slower
(smaller $r_{i}(t,t+\Delta t)$) than initially less indebted economies.  Hence, $\beta$ represents the ``speed of
convergence'', a concept  introduced for {\it per capita\/} GDP data in Ref. \cite{martin96}.  A larger
positive 
value of $\beta$ results in faster convergence, equalizing the {\it per capita\/} debt across all
countries more quickly. If $\beta<0$ there is divergence in debt data, where initially less-indebted
countries with smaller $d_i(t)$ increase their debt slower than initially more-indebted countries.

In Fig.~\ref{figure1} we plot the 1990 {\it per capita\/} debt of more than 80 countries, representing  low-, medium-,
and high-wealth countries, considering several relationships.  First, we compare the {\it per capita\/} debt over (a)  15-year
and (d)  7-year  time horizons. We find the slope $S= (1-\beta\Delta t)$ of the regression in
Eq.~(\ref{loglogregress}) is less than one, requiring $\beta >0$ which  corresponds to scenario (iii). 

To confirm the convergence across countries for other time horizons,  Fig.~\ref{figure2} shows the
value of $S$ for varying initial $d(t)$ and time horizon $\Delta t$ in
 Eq.(\ref{loglogregress}).  We find  $S \equiv 1-\beta\Delta t <1$ for most horizons $\Delta t$, implying 
convergence, where less-indebted countries increase their debt faster than more-indebted countries. However, 
there is a period in the beginning of the 1990's that is the exception, with $S>1$ and $\beta <0$. This period of
divergence in {\it 
per capita} debt may be related to the 20-year lows in interest rates which may have resulted in increased borrowing,
even among 
heavily indebted countries. Since 1995, the values of $S$ have returned to values less than one, indicating a return to
convergence.

We also analyze the growth rates of {\it per capita} GDP and confirm the divergence across countries observed
originally
in \cite{barromartin91, barromartin92,  martin96}. In Fig. \ref{figure8} we plot  for {\it per capita} GDP, the
analogous regression $S$ values that we plot in Fig. \ref{figure2} for {\it per capita} public debt. For all periods
$\Delta t$ and initial years analyzed,  we find  values of $S>1$ indicating divergence.

A natural question is  -- how does the {\it per capita} debt $d_{i}$ vary across all countries and by income group?
Power law probability density functions (pdf) have been  observed for total country GDP \cite{globalization2} and {\it per capita} GDP
\cite{GDPpercapPDF}. Fig. \ref{debtPDF} shows the pdf  $P(d)$
 for all countries analyzed over the 36-year period 1970-2005. 
 We observe large variations across income groups, where low income countries typically have relatively small {\it per
capita} debt values reflecting their small per capita borrowing capacity. In contrast to the zipf-rank curves for
GDP with $\zeta_{GDP} \approx 1$ corresponding to pdf scaling exponent exponent $1+1/\zeta_{GDP} \approx
2$\cite{globalization2}, we
observe in Fig. \ref{debtPDF}(inset) a scaling value $\zeta_{d} \approx 0.3$ corresponding to a relatively large pdf scaling exponent $1+1/\zeta_{d} \approx 4.3$.

In a country where both GDP and debt grow with time, one must analyze the dynamics of both debt and GDP. 
Since a debt that is large for Luxembourg is not large for the U.S., various indices have been proposed in order to
compare the burden of debt to the ability of the country's economy to generate income.
These include ${\cal R} $\cite{blans89, ecogrowth}, so we apply the
convergence analysis of Eq.~(\ref{loglogregress}) to ${\cal R}(t)$ obtaining
\begin{equation}
\log[{\cal R}_{i}(t+\Delta t)]= \alpha' ~\Delta t +(1-\beta'\Delta t) \log[{\cal R}_{i}(t)] \ .
\label{loglogregress2}
\end{equation}
Fig. \ref{figure1} compares ${\cal R}(t)$ over (b) 15-year and (e) 7-year  time
horizons.  
Fig. \ref{figure4} shows $S'  \equiv (1-\beta'\Delta t) < 1$, implying convergence $ \beta' >0$, over a large range of
$\Delta t$-year horizons for  initial year $t$. 

A responsible government is expected to monitor simultaneously the growth of debt and GDP \cite{blanchard94}. By 
borrowing money, a country may increase ${\cal R}(t)$ for some time,  but clearly ${\cal R}(t)$
cannot increase indefinitely, as increased debt can negatively affect GDP growth \cite{saint}. Banks prefer individuals
with large incomes 
and small debts. Banks also prefer countries that have, for a given  GDP level, small relative debt.  Fig. \ref{DbyGExamples}  provides
 the annual trend of ${\cal R}(t)$ for several groups of countries with common geo-politial backgrounds.

   Debtor default risk is estimated by many rating agencies and
   financial  organizations. ${\cal R}(t)$ is an important quantity for determining the  ability of a debtor to
make debt payments. For large ${\cal R}(t)$ there is a
   larger  probability that the debtor will not be able to  make timely payments or be able to prevent further debt increase
with time, scenarios that lead to credit default. In order to quantify the risky debt levels, 
we collect the ${\cal R}(t)$ values of all countries analyzed over the 36-year period 1970-2005 and plot the
pdf $P({\cal R})$ in   Fig. \ref{DbyGPDF}. We find $<{\cal R}> = 0.57 \pm 0.54$, and we 
fit the pdf to a Gamma distribution $P({\cal R}) \propto {\cal R}^{k-1} \exp [-{\cal R}/{\cal R}_{c}]$ with
 $k=2.0 \pm 0.1$ and ${\cal R}_{c}= 0.30 \pm 0.01$, using the maximum likelihood estimator.  The extreme value statistics 
 of Gamma pdf can be used to define thresholds for sustainable debt.

In order to analyze the countries that   have large ${\cal R}(t)$ and a high risk of
default, the countries which constitute the pdf tail,  we plot  rank-frequency curves in Fig. \ref{DbyGPDF}(inset). The Zipf
plots show  a power law over three orders of magnitude, with scaling
exponent $\zeta_{\cal R} \approx 0.4$ corresponding to $P({\cal R}) \sim {\cal R}^{-3.5}$.

\section{Model}
Our analysis, performed across a wide cross section of countries, confirms the existence of convergence in  public 
debt. This is opposite of what is in  GDP data \cite{barromartin91, barromartin92,  martin96}, where the speed of convergence $\beta$ is negative. 
We now discuss how to model  the scaling result we obtain, 
 and  how to use the scaling result obtained for GDP and public debt. 
Figs.~\ref{figure1}(c) and \ref{figure1}(f)  compare the {\it per capita\/} debt to {\it per capita\/} GDP for the years 1990 and 2005. 
The typical relationship between debt and GDP shows a  scale invariant form, 
\begin{eqnarray}
g \sim A~d^{\gamma} \ ,
\label{GD}
\end{eqnarray}
where $g$ is the {\it per capita\/} GDP and $d$ is the {\it per
capita\/} debt. In Fig. \ref{figure4} we plot the values of  $\gamma$ for the set of countries analyzed in  each
yearly data set. 

 In order to model debt dynamics, we assume that the functional dependence in Eq.~(\ref{GD}) is time invariant.
  Note that if $g(t)$ and $d(t)$ grow exponentially with different growth rates, $r_{g}$ and $r_{d}$, the relationship
between $g(t)$ and $d(t)$ still has the form of a power law, with  $r_{g} = \gamma \ r_{d}$. 

We may consider the dynamics of public debt by assuming that the government borrows $B (t)$, a fixed proportion of GDP given by $B (t) \equiv
D(t)-D(t-1) \equiv \Delta D(t) = b ~ G(t)$, 
with deficit ratio $b= $ constant \cite{brauninger}.  Then $r_{D} \equiv
\Delta D / D = b ~ G(t) /D(t) $, and 
\begin{equation}
r_{d} = b ~g(t) / d(t) - r_{pop}  = bN / d(t)^{1-\gamma} - r_{pop}\ ,
\end{equation} 
where $r_{pop}$ denotes the population growth rate \cite{ecogrowth}. Hence, 
\begin{equation}  
\frac{
\partial r_{d}}{ 
\partial d} = \frac{-b N(1-\gamma)}{d ^{2 - \gamma}} \propto 1/d ^{2-\gamma} \ .
\end{equation} 
We observe from Fig. \ref{figure4}  that $\gamma \lesssim 1$ so that  $\partial r_{d} / 
\partial d \simeq -1/d$. For this reason, we use the regression  $r_{d}  \cong \alpha-\beta ~\log[d_i(t)]$, which
agrees well with the data in Figs. (\ref{figure1})  and  (\ref{figure2}).

\section{Discussion}

In summary, we demonstrate {\it convergence} in {\it per capita} public debt across a wide set of countries during the period
1970-2005, a result of general interest for complex systems researchers as well as for creditors. 
Our analysis  is made possible by new comprehensive data sets, and extends  empirical surveys 
previously performed on country GDP which found {\it divergence} in country GDP. While divergence in country GDP implies
that 
economic wealth is moving away from global equality, convergence in {\it per capita} debt implies that indebtedness is
becoming an economic 
standard. 
Furthermore, convergence in the ${\cal R}$ implies that relative differences in indebtedness across countries is 
also decreasing over time. 
Some economists believe that convergence across all countries is possible through  globalization 
\cite{globalization1,globalization2,globalization3,globalization4} and access to open markets \cite{SachsEconReform}.
While public debt can be used to invest in a country's development via physical infrastructure, technology, and social
programs, its use requires responsible 
governance. Corruption\cite{Corruption1,Corruption2} and the misuse of public debt can lead to insurmountable debt
contributing to financial crisis, which can  cause further increase in debt levels 
through exchange rate depreciation \cite{bolle,sachs88}. 
There are also instances of extremely poor debtor nations that are unable to meet their current and future debt
obligations (See Fig. \ref{DbyGExamples}). Recent programs such as the Heavily Indebted Poor Countries (HIPC)
Initiative, sponsored by the World Bank and the IMF, and the Jubilee 2000 Campaign-to-Drop-the-Dept, have called on
debt cancellation for extremely poor debtor nations as a crucial step in the UN Millennium Project to eliminate extreme
poverty \cite{SachsPoverty, SachsPovertyAfr, UNMilleniumProject}. Further, debt has become a 
problem for not only the extremely poor countries. With the current global credit crunch, their is an increased need for
responsible use of  government debt.

\bigskip\bigskip

\acknowledgments We thank  Laurence J. Kotlikoff, S. N. Durlauf, and Ivo Krznar for helpful discussions, and NSF 
for support.

 \clearpage
 \newpage
 
 \begin{widetext} 
 
   \begin{figure}
\centering{\includegraphics*[width=0.8\textwidth]{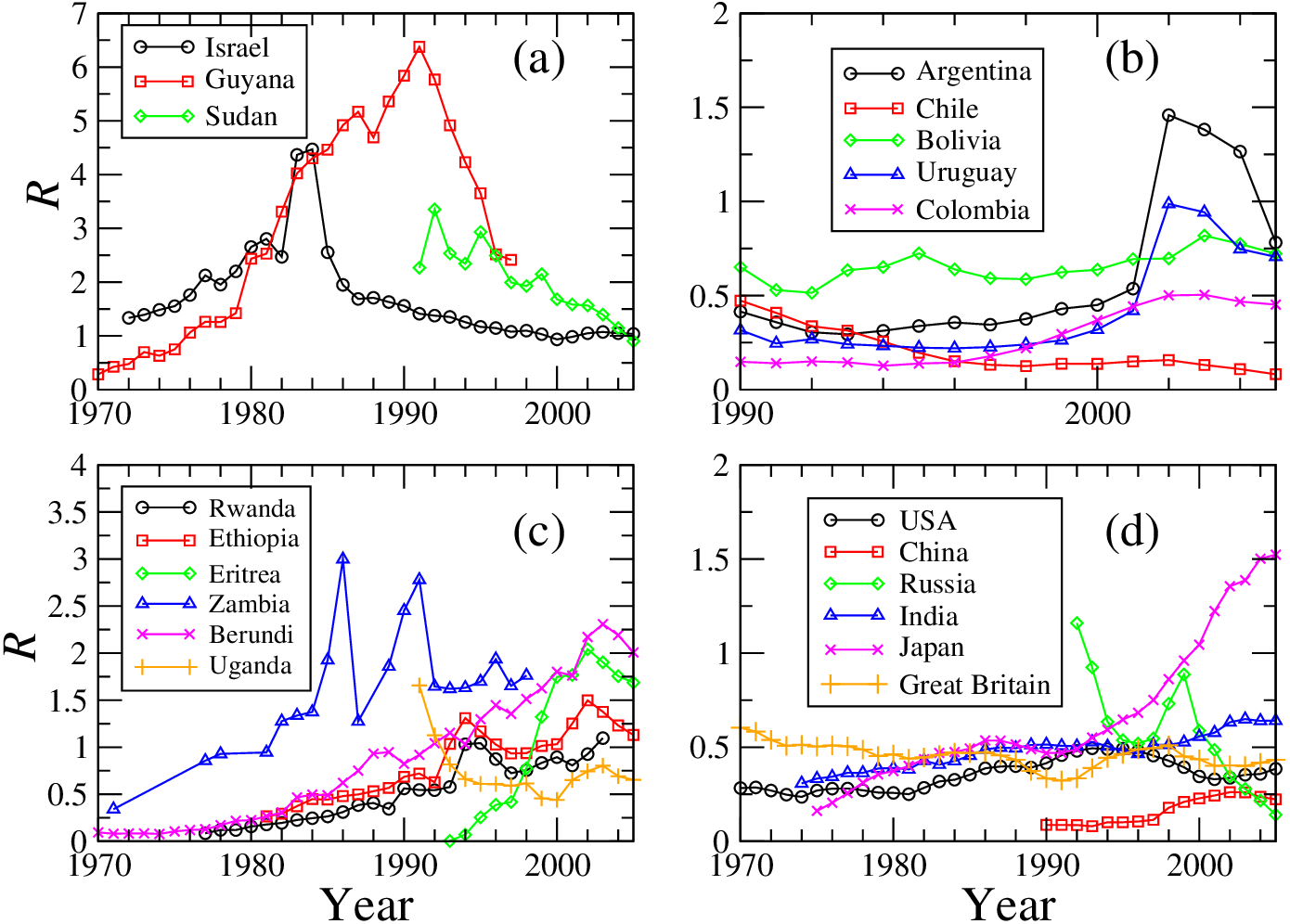}}
\centering{\includegraphics*[width=0.8\textwidth]{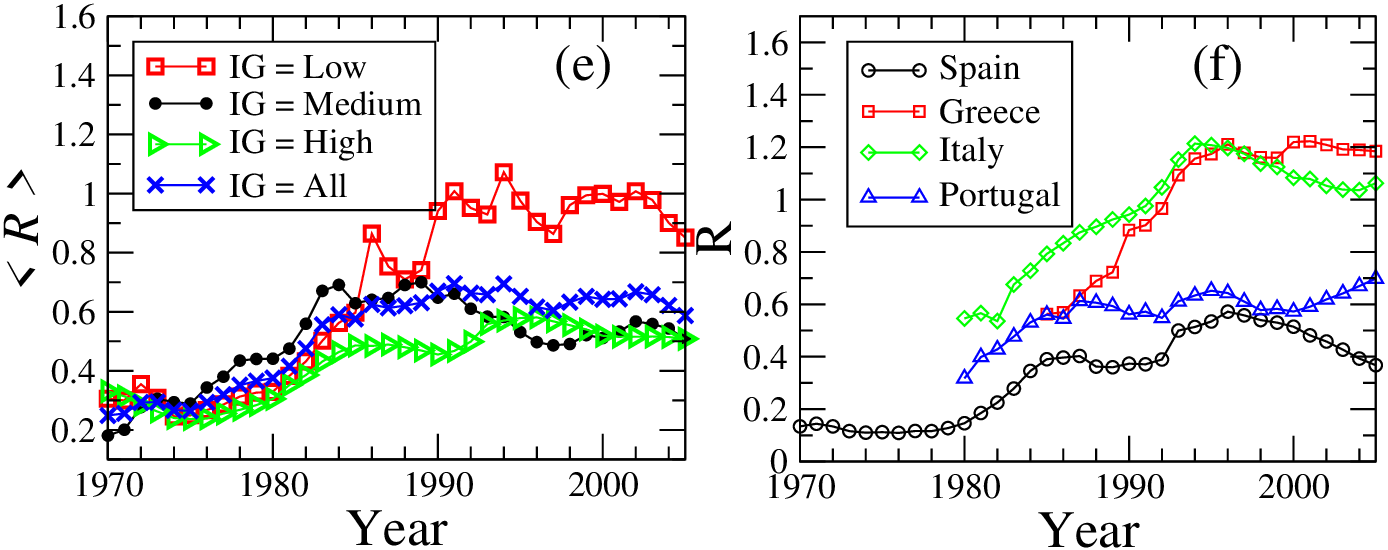}}
  \caption{ \label{DbyGExamples}   Illustration of debt-to-GDP ratio ${\cal R}(t)$ for (a-d) several countries and (e)
the average trend for three subsets according to World Bank Income Group (IG) classifications. (a) Countries with
turmoil, as in the case of Israel and Sudan which experienced periods of war,  and Guyana which formed a new government,
can experience periods of extremely high ${\cal R}$. (b) The Argentine financial crisis of the late 1990's placed many
South American countries in a state of debt stress. (c) The debt burden in many African countries, some which are also
plagued by war and political turmoil, is increasing with time. Many of these countries are candidates for foreign debt
cancellation (jubilee). (d) Several large countries. (e)  The average ${\cal R}$  is increasing with
time, with low-income countries currently at relatively high levels reflecting the burden of  unsustainable debt. (f) Mediterranian countries with recent concern of default risk.}
\end{figure}

  \begin{figure}
\centering{\includegraphics*[width=0.65\textwidth]{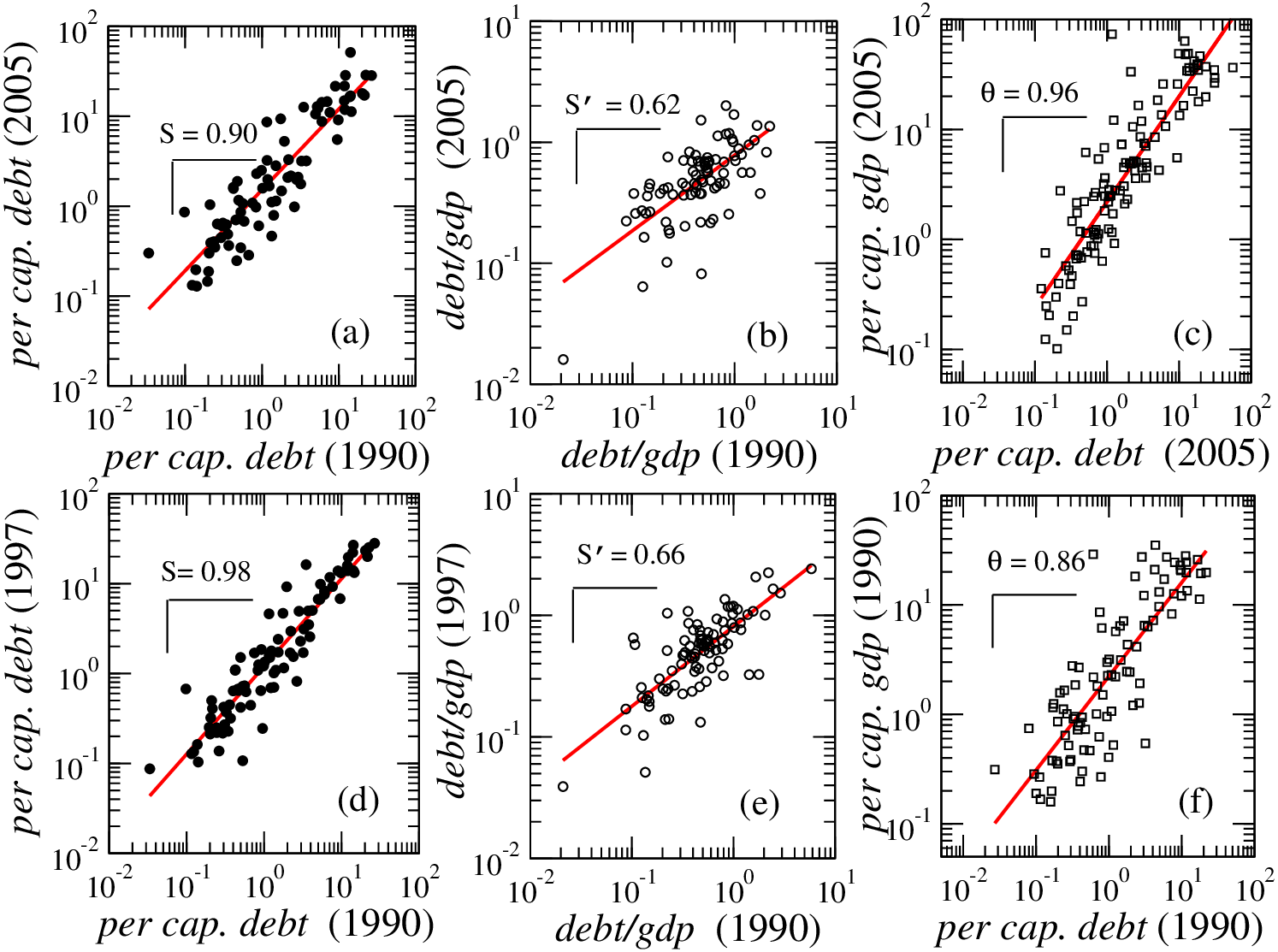}}
  \caption{ \label{figure1} The 3 possible relations between {\it per capita\/} debt,  {\it per capita\/} GDP, and ${\cal R}$ for a set of low-income, medium-income, and high-income countries. Regression of {\it per
capita\/} debt $d(t)$ in 1990 versus  {\it per capita\/} debt $d(t + \Delta t)$ for (a) $\Delta t = 15$ years  and  (d) $\Delta t = 7$ years later indicates that there is {\it
convergence} in the {\it per
  capita\/} debt data across the countries over the larger time horizon in panel (a) with $S \equiv 1-\beta\Delta t<1$ implying $\beta >0$.
Regression of ${\cal R}(t)$ in 1990 versus  ${\cal R}(t + \Delta t)$ for (b) $\Delta t =$ 15 years  and  (e) $\Delta t =$ 7 years indicates
that there is also {\it convergence} over these time horizons. (c,f) We also illustrate the  scale invariant relation between between
{\it per capita\/} debt and {\it per capita\/} GDP in units of $10^{3}$ {\it USD} per person in the year 2000.  }
\end{figure}

 \begin{figure}
\centering{\includegraphics*[width=0.65\textwidth]{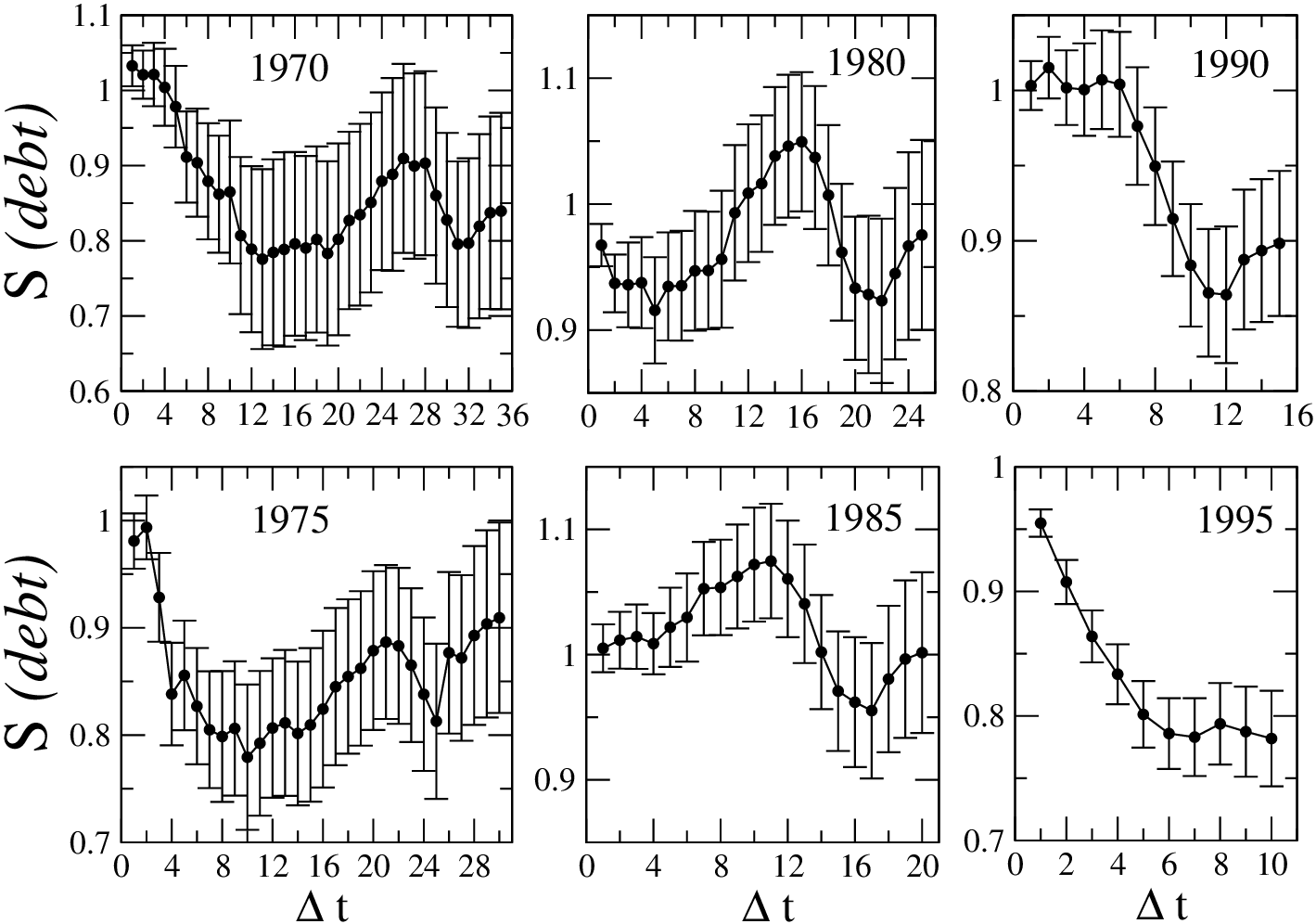}}
  \caption{ \label{figure2}The slope $S \equiv  1 - \beta \Delta t $ of the log-log regression in Eq.~(\ref{loglogregress}) is typically less than
one for a given time horizon of $\Delta t$ years after the initial year $t$ labeling each panel. Values of $S<1$
imply {\it convergence} ($\beta >0$) in {\it per capita\/} debt over the $\Delta t$-year period. $R^{2}$ values for
log-log regressions are typically $>$ 0.7 for $10 < \Delta t < 20$ years, and are typically  $>$ 0.8 for $1 < \Delta t <
10$ years, while the number of countries for each regression varies between 22-25 (1970) and 103-120 (1995).
 Interestingly, we observe a short period of {\it divergence}, where for initial years 1980 and 1985,
the value of $S>1$ for  $\Delta t \approx 12$ years corresponding to the mid-1990's. }
\end{figure}

 \begin{figure}
\centering{\includegraphics*[width=0.65\textwidth]{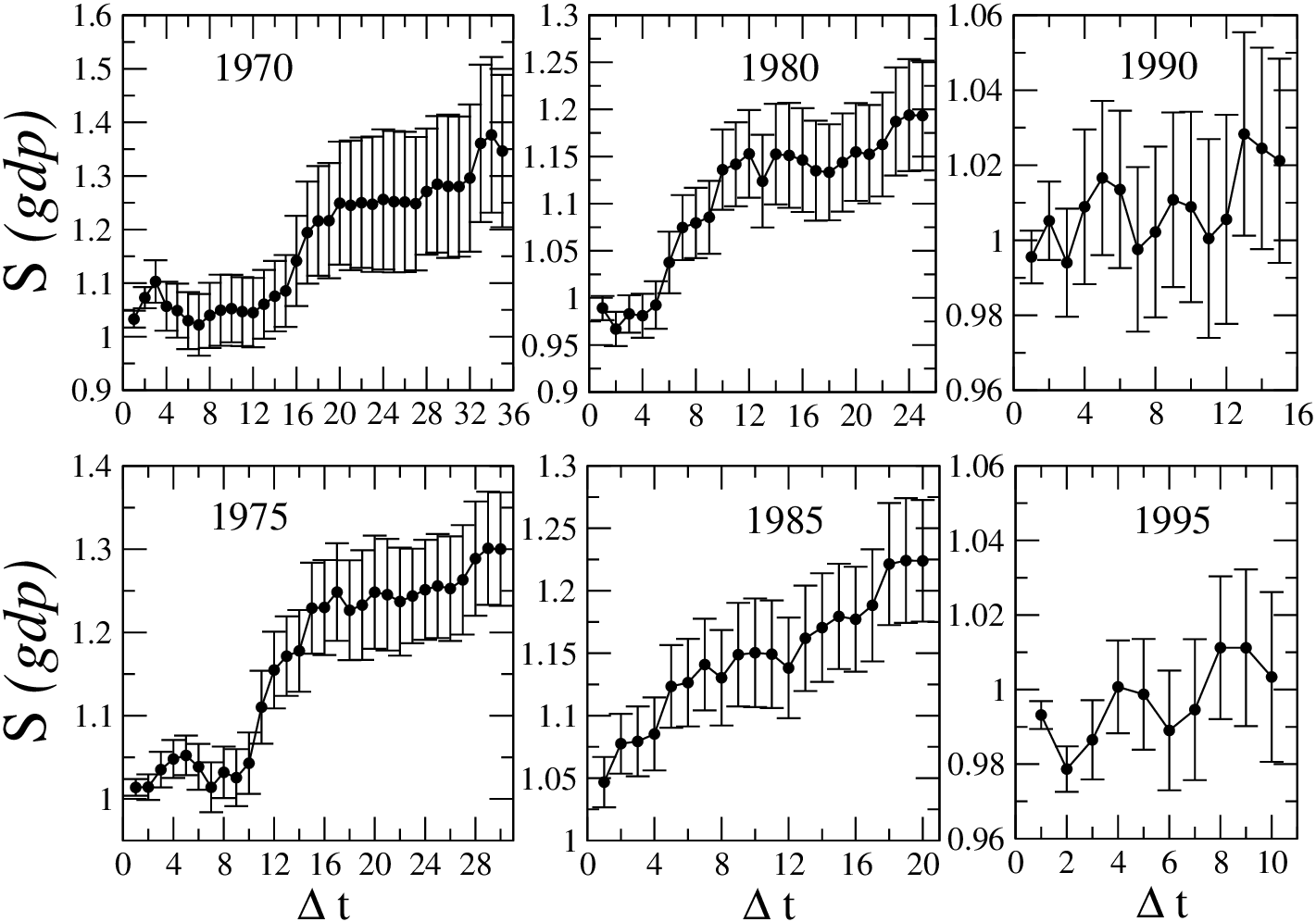}}
 \caption{ \label{figure8} For comparison, we also analyze  GDP growth rates and find  {\it divergence} for all
periods analyzed.
  The slope $S$ of the log-log regression in Eq.~(\ref{loglogregress}) for {\it per capita} GDP is typically {\it
greater} 
than
one for a given time horizon of $\Delta t$ years after the initial year, which we label in each panel. Values of $S>1$
imply {\it divergence} ($\beta <0$) in {\it per capita\/} GDP over the $\Delta t$-year period, consistent with the
seminal 
work of \cite{barro91,martin96}.}
\end{figure}

\begin{figure}
\centering{\includegraphics*[width=0.65\textwidth]{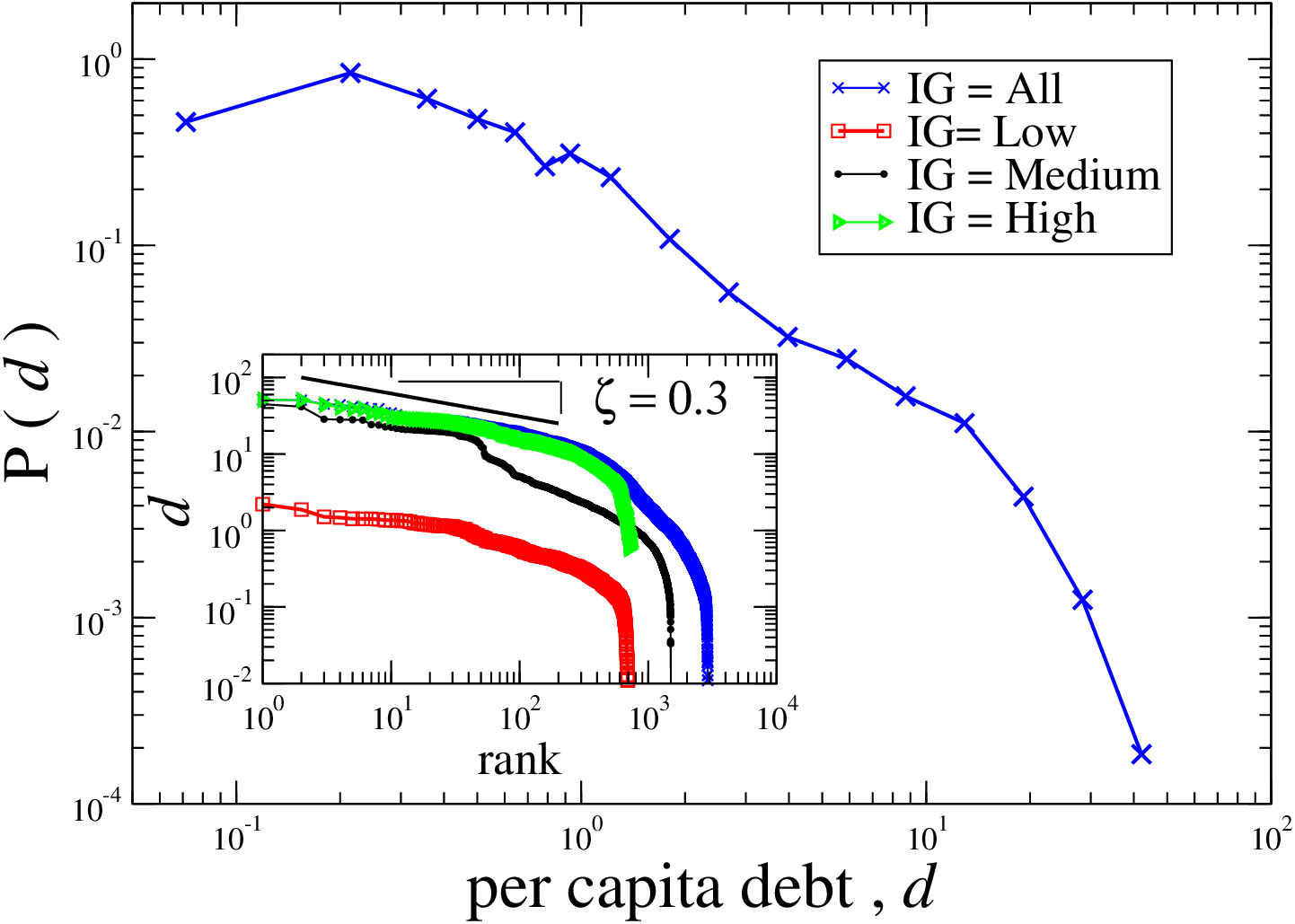}}
  \caption{ \label{debtPDF} Right-skewed probability density function $P(d)$ of  {\it per capita} debt $d$ for all
countries analyzed over the 36-year period 1970-2005  in units of $10^{3}$ {\it USD} per person in the year 2000. (Inset) Zipf
rank-frequency plot for $d$ values
  for all countries and for three subgroups according to {\it World Bank} Income Group (IG) classification.  Countries in the
high and medium IG categories contribute 
  to the wide range of $d$ values, whereas poor countries have relatively small $d$ values reflecting their small per
capita borrowing capacity. We also show a power-law with exponent $\zeta = 0.3$ for comparison. }
\end{figure}

  \begin{figure}
\centering{\includegraphics*[width=0.8\textwidth]{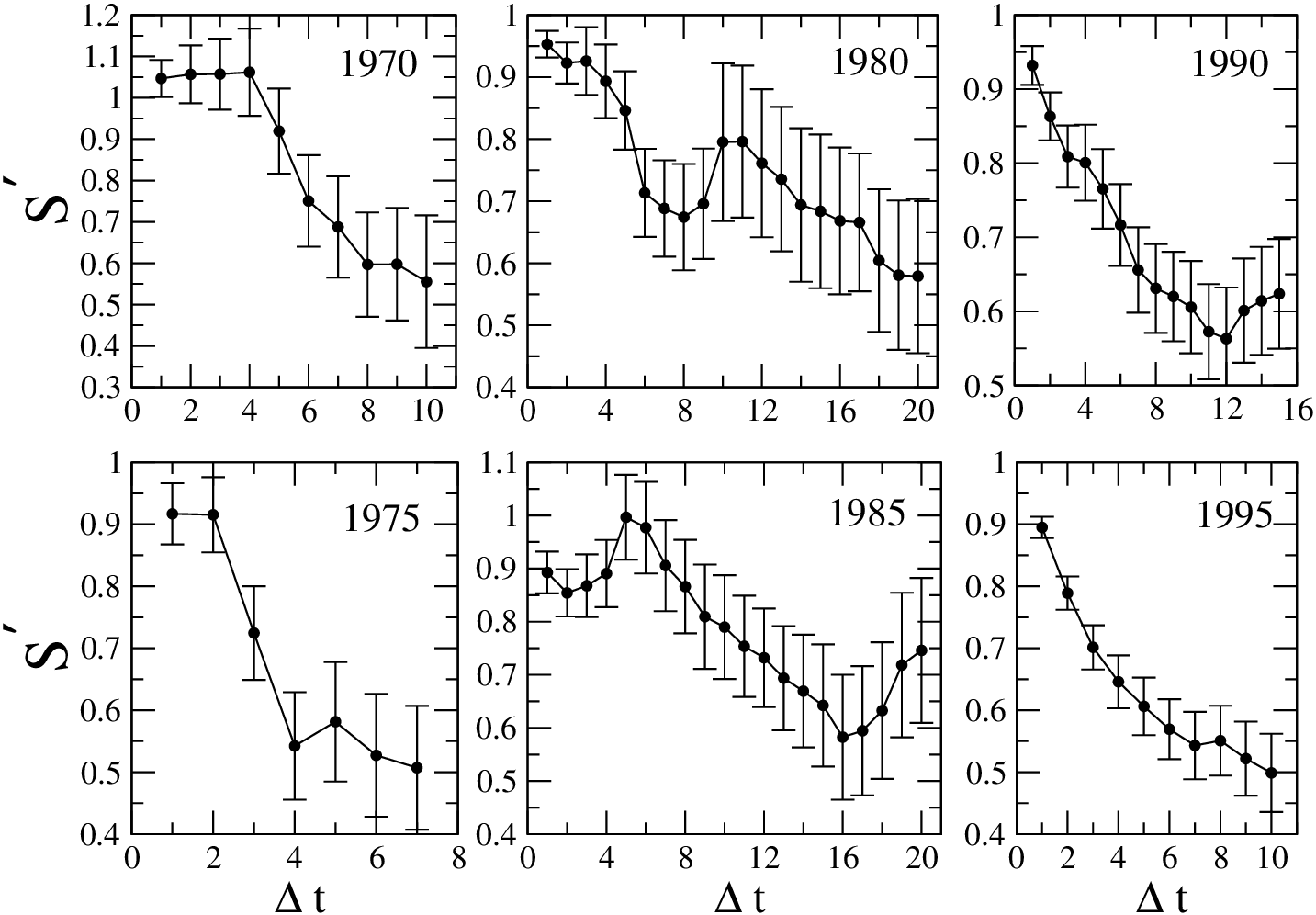}}
  \caption{ \label{figure4}  The slope $S' \equiv  1-\beta'\Delta t $ of the log-log regression in Eq.~(\ref{loglogregress2})
of  ${\cal R}$. We find  $S' < 1$ a given time horizon of $\Delta t$ years after the
initial year $t$ labeling each panel. Values of $S<1$ imply {\it convergence} ($\beta >0$) in {\it per capita\/}
debt over the $\Delta t$-year period. We plot values of $S'$ only for log-log regressions with $R^{2} > 0.3$. The number
of countries for each regression varies between 22-25 (1970) and 103-120 (1995).}
\end{figure}

   \begin{figure}
\centering{\includegraphics*[width=0.8\textwidth]{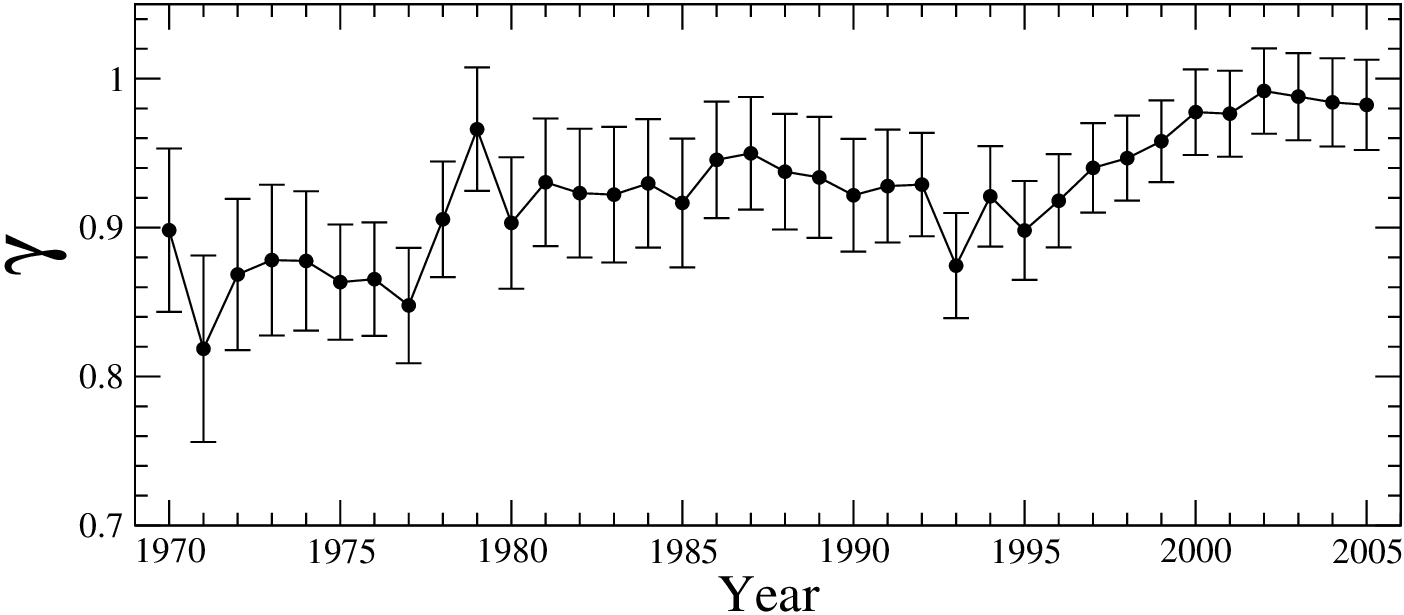}}
\caption{ \label{figure4}  Annual trend of  power-law exponent $\gamma$,  which quantifies
the scaling relationship between debt and  GDP as $g \sim A \ d^{\gamma}$ in Eq.~(\ref{GD}). 
We find $\gamma < 1$ which reflects the burden of debt on country GDP. 
  }
\end{figure} 

 \begin{figure}
\centering{\includegraphics*[width=0.8\textwidth]{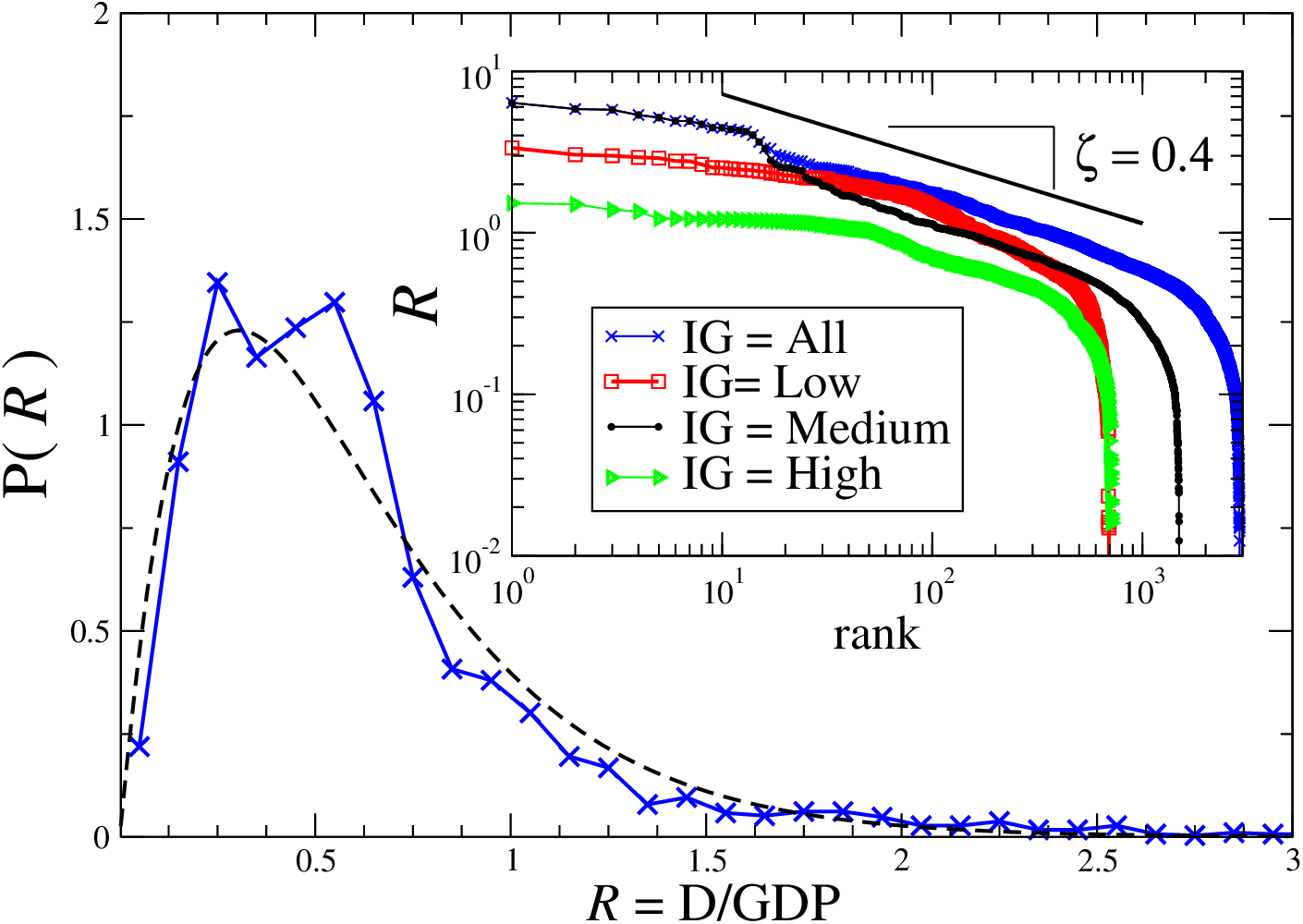}}
  \caption{ \label{DbyGPDF} Probability density function $P({\cal R})$  for all countries analyzed over the 36-year period 1970-2005.
  For comparison, we approximate $P({\cal R})$ with a Gamma distribution with parameters $k=2.0$ and ${\cal R}_{c} =
0.30$. The extreme statistics of $P({\cal R})$ can quantify the threshold of sustainable debt. (Inset) Zipf rank-frequency plot for ${\cal R}$ values. We
  separate data into three subgroups by {\it World Bank} Income Group (IG) classification to demonstrate the range of large
${\cal R}$ values within different wealth groups. The straight line is a power law with exponent $\zeta = 0.4$ for comparison.  }
\end{figure}

\end{widetext}

\end{document}